\input harvmac.tex
\input amssym.tex

\cal{}

\font\cmss=cmss10
\font\cmsss=cmss10 at 7pt

\def\IL{\relax{\rm I\kern-.18em L}}
\def\IH{\relax{\rm I\kern-.18em H}}
\def\IR{\relax{\rm I\kern-.18em R}}
\def\inbar{\vrule height1.5ex width.4pt depth0pt}
\def\IC{\relax\hbox{$\inbar\kern-.3em{\rm C}$}}
\def\rlx{\relax\leavevmode}

\def\ZZ{\rlx\leavevmode\ifmmode\mathchoice{\hbox{\cmss Z\kern-.4em Z}}
 {\hbox{\cmss Z\kern-.4em Z}}{\lower.9pt\hbox{\cmsss Z\kern-.36em Z}}
 {\lower1.2pt\hbox{\cmsss Z\kern-.36em Z}}\else{\cmss Z\kern-.4em
 Z}\fi}
\def\IZ{\relax\ifmmode\mathchoice
{\hbox{\cmss Z\kern-.4em Z}}{\hbox{\cmss Z\kern-.4em Z}}
{\lower.9pt\hbox{\cmsss Z\kern-.4em Z}}
{\lower1.2pt\hbox{\cmsss Z\kern-.4em Z}}\else{\cmss Z\kern-.4em
Z}\fi}


\font\manual=manfnt
\def\dbend{\lower3.5pt\hbox{\manual\char127}}

\def\IZ{\relax\ifmmode\mathchoice
{\hbox{\cmss Z\kern-.4em Z}}{\hbox{\cmss Z\kern-.4em Z}}
{\lower.9pt\hbox{\cmsss Z\kern-.4em Z}} {\lower1.2pt\hbox{\cmsss
Z\kern-.4em Z}}\else{\cmss Z\kern-.4em Z}\fi}


\def\rt2{\sqrt{2}}
\def\irt2{{1\over\sqrt{2}}}


\Title{\vbox{\hbox{DISTA-2011}
\hbox{DFTT 26/2011}}} 
{\vbox{ 
\centerline{BCJ and KK Relations from BRST Symmetry
} \vskip .2cm
\centerline{and Supergravity Amplitudes}}}

\medskip
\centerline{P.~A. Grassi$^{~a,b,}$\foot{pgrassi@cern.ch}, A.~Mezzalira$^{~c,b,}$\foot{mezzalir@to.infn.it},
and L. Sommovigo$^{~a,}$\foot{lsommovi@mfn.unipmn.it} }
\medskip 
\centerline{${a)}$ {\it DISTA, Universit\`a del Piemonte Orientale,} }
\centerline{\it via T. Michel, 11, 15121 Alessandria, ITALY.}
\medskip
\centerline{${b)}$ {\it INFN, Gruppo Collegato di Alessandria, Sezione di Torino, ITALY}}
\medskip
\centerline{${c)}$ {\it Dipartimento di Fisica Teorica, v. P. Giuria,1, 10125, Torino, ITALY}}
\medskip
\vskip  .5cm
We use Pure Spinor string theory to construct suitable kinematical factors which 
explicitly satisfy  the Kleiss-Kuijf (KK) relations.
Using the formula conceived by Bern et al. and employed by us for 4- and 5-point amplitudes in a previous work, we are able to 
compute the 6-point supergravity amplitude from the corresponding 
SYM building blocks given by Mafra et al.. 
 We derive the KK and Bern-Carrasco-Johansson (BCJ) 
identities from the BRST invariance and we discuss the relations between Bern et al. building
blocks and those of Mafra et al..

\Date{\ September 2011}

\lref\MafraJQ{
  C.~R.~Mafra, O.~Schlotterer, S.~Stieberger and D.~Tsimpis,
  arXiv:1012.3981 [hep-th].
}
\lref\MafraGJ{
  C.~R.~Mafra, O.~Schlotterer, S.~Stieberger, D.~Tsimpis,
Nucl.\ Phys.\  {\bf B846}, 359-393 (2011).
[arXiv:1011.0994 [hep-th]].
}

\lref\MafraWQ{
  C.~R.~Mafra,
  arXiv:0902.1552 [hep-th].
}

\lref\BernIA{
  Z.~Bern and T.~Dennen,
  arXiv:1103.0312 [hep-th].
}

\lref\MafraIR{
  C.~R.~Mafra,
  JHEP {\bf 1011}, 096 (2010)
  [arXiv:1007.3639 [hep-th]].
}

\lref\CarrascoMN{
  J.~J.~M.~Carrasco, H.~Johansson,
[arXiv:1106.4711 [hep-th]].
}
\lref\BernFY{
  Z.~Bern, J.~J.~M.~Carrasco, H.~Johansson,
Nucl.\ Phys.\ Proc.\ Suppl.\  {\bf 205-206}, 54-60 (2010).
[arXiv:1007.4297 [hep-th]].
}
\lref\MafraPN{
  C.~R.~Mafra,
[arXiv:1007.4999 [hep-th]].
}
\lref\MafraKJ{
  C.~R.~Mafra, O.~Schlotterer, S.~Stieberger,
[arXiv:1104.5224 [hep-th]].
}
\lref\MafraNV{
  C.~R.~Mafra, O.~Schlotterer, S.~Stieberger,
[arXiv:1106.2645 [hep-th]].
}
\lref\MafraNW{
  C.~R.~Mafra, O.~Schlotterer, S.~Stieberger,
[arXiv:1106.2646 [hep-th]].
}

\lref\BernYG{
  Z.~Bern, T.~Dennen, Y.~-t.~Huang, M.~Kiermaier,
Phys.\ Rev.\  {\bf D82}, 065003 (2010).
[arXiv:1004.0693 [hep-th]].
}
\lref\BernQJ{
  Z.~Bern, J.~J.~M.~Carrasco, H.~Johansson,
Phys.\ Rev.\  {\bf D78}, 085011 (2008). 
[arXiv:0805. 3993 [hep-ph]].
}
\lref\BernIA{
  Z.~Bern, T.~Dennen,
[arXiv:1103.0312 [hep-th]].
}
\lref\BjerrumBohrRD{
  N.~E.~J.~Bjerrum-Bohr, P.~H.~Damgaard, P.~Vanhove,
PRLTA,103,161602.\ 2009 {\bf 103}, 161602 (2009).
[arXiv:0907.1425 [hep-th]].
}
\lref\KleissNE{
  R.~Kleiss, H.~Kuijf,
Nucl.\ Phys.\  {\bf B312}, 616 (1989).
}
\lref\BianchiPU{
  M.~Bianchi, H.~Elvang, D.~Z.~Freedman,
JHEP {\bf 0809}, 063 (2008).
[arXiv:0805.0757 [hep-th]].
}
\lref\MaUM{
  Q.~Ma, Y.~-J.~Du, Y.~-X.~Chen,
[arXiv:1109.0685 [hep-th]].
}

\lref\GrassiUF{
  P.~A.~Grassi, L.~Sommovigo,
[arXiv:1107.3923 [hep-th]].
}

\lref\BjerrumBohrZS{
  N.~E.~J.~Bjerrum-Bohr, P.~H.~Damgaard, T.~Sondergaard, P.~Vanhove,
JHEP {\bf 1006}, 003 (2010).
[arXiv:1003.2403 [hep-th]].
}

\lref\BjerrumBohrRD{
  N.~E.~J.~Bjerrum-Bohr, P.~H.~Damgaard, P.~Vanhove,
Phys.\ Rev.\ Lett.\  {\bf 103}, 161602 (2009).
[arXiv:0907.1425 [hep-th]].
}


\newsec{Introduction}

In the case of open strings, Mafra et al. \refs{\MafraIR,\MafraJQ} constructed a set of building blocks 
from Pure Spinor string theory with nice and interesting properties. They are classified 
in terms of their BRST transformation rules, ghost numbers and the number of the external legs.
In terms of these building blocks the authors of \MafraKJ\ are able to describe n-point amplitudes for SYM. These results can be achieved directly from the limit of string theory \refs{\MafraWQ,\MafraGJ,
\MafraPN}
or from SYM Feynman diagrams computations \refs{\CarrascoMN,\BernFY}.

Furthermore, a convenient decomposition of  4-, 5- and 6-point in SYM singles out some expressions (kinematical factors), 
denoted by ``$n$'', satisfying very useful algebraic relations known as BCJ \refs{\BernQJ}. 
These relations have been derived from string theory by worldsheet methods (monodromy \refs{\BjerrumBohrZS,\BjerrumBohrRD,\MaUM}) and
directly confirmed by SYM amplitude computations \BernQJ.
Finally, it is certainly convenient to implement those relations in terms of Mafra's building blocks.
This has been done in \MafraKJ\ where, using again worldsheet computations
of multi-leg massless external-state amplitudes, the appropriate combinations of those blocks
have been discovered. 

Starting from a different approach, namely from direct computation of SYM amplitudes, some new building blocks, named ``$\tau$'', have been described in paper \BernIA.
They satisfy new relations known as KK, discovered for QCD amplitudes in \KleissNE.
The role of KK for $\tau$ and BCJ for $n$ is to reduce the number of independent building blocks
to the correct number which, for a tree level p-point amplitude, is $\left( {\rm p}-2 \right)!$. When one expresses 
the amplitudes in terms of building blocks fulfilling the KK relations, they turn out to obey also the BCJ 
relations (and consequently, the number 
of independent amplitudes is $({\rm p}-3)!$). On the other side, if the building blocks obey the BCJ relations, the corresponding amplitudes 
satisfy the KK relations. 
It has also been pointed out by 
Bern et al., in the analysis of SYM amplitudes, that KK relations play a fundamental role 
in the finiteness of the theory and in the cancellations due to supersymmetry \BernYG. 

Furthermore, Mafra et al. showed that some BCJ relations can be derived from BRST symmetry 
and, when the amplitudes are written in terms of those ingredients, they display the wanted 
symmetry properties. Here, we show that indeed all possible BCJ relations 
are found by using BRST symmetry. We explicitly check it on  the 4-, 5- and 6-point amplitudes.

In the same paper \BernIA, Bern et al. proposed a new formula for supergravity amplitudes based on the SYM building blocks, which has been used by two of the authors of the present paper to derive the supergravity amplitudes for 4- and 5-point \GrassiUF. 
That formula uses the $\tau$'s as follows: first they rewrite the open string amplitude
$A^{Open}\left( n \right)$ in terms of $\tau_L$'s by means of a relation between $n$'s and $\tau_L$'s, that is $A^{Open}\left( n \right)\rightarrow A^{Open}\left( n\left( \tau_L \right) \right)=A_{L}\left( \tau_L \right)$. Then this expression (thought as the left-mover part of closed string amplitudes) is 
paired up with the right-mover $\tau_R$ as $A_{L}\left( \tau_{L} \right) \tau_{R}$
and summed over the non-equivalent permutations of the external legs.

To check if this procedure is correct, we adopt Mafra's point-of-view: the construction 
is based on the decomposition in terms of the building blocks and it has to respect the 
BRST symmetry.\foot{As a matter of clearness, we underline that for BRST invariance of 
an amplitude, we mean the BRST invariance of the integrand before taking the projection onto physical states.} 
Therefore,  
we require the supergravity amplitude to be invariant under the 
left- and right-BRST symmetry. This can easily be checked, observing that the  
SYM amplitude is manifestly invariant under the left-BRST variations, which  leave invariant 
the $\tau_R$'s, used to construct the supergravity amplitudes. Vice-versa, the same is true if the specular construction, namely by taking the $\tau_L$ and $A(\tau_R)$, can be considered. 
Therefore, the main issue is to prove the equivalence of the two expressions or, differently said, to prove the expression for 
the amplitude to be symmetric under the exchange of chiralities.  
In the case of 4- and 5-point functions, this fact has been shown in \GrassiUF, but in the case of 6-point functions it turns out to be rather more difficult. 
Here, we review the computation of \GrassiUF\ from a different perspective, we explicitly show 
that there is a suitable combination of Mafra's building blocks -- which coincide with Bern's $\tau$'s -- that 
renders the supergravity amplitude symmetric in the exchange of chiralities. For the 6-point functions, we construct such a combination and show that our solution enjoys the needed symmetry requirements. 

The paper is organized as follows: in sec. 2, we briefly review the BCJ and KK relations.  
We show that all BCJ relations can be derived from the BRST symmetry. In particular, it is proven that 
the BRST exactness of some linear combinations of vertex operators -- which, in turn, implies the vanishing of 
the corresponding amplitude -- are directly linked to BCJ relations. In sec. 3, we discuss the relation between 
different types of building blocks, their symmetry properties and the relations they must fulfill. Finally, in sec. 4 
we discuss the 6-point functions. In appendix A we present a compact form for the BRST transformations of the building blocks. 


\newsec{BCJ, KK relations and $Q$-exactness}

Both KK and BCJ are induced by monodromy as shown in \refs{\BjerrumBohrRD,\BjerrumBohrZS}. 
In order to discuss this argument, let us consider a tree level n-point open string amplitude with $U\left( N \right)$ gauge group
\eqn\stringopenamplgeneral{
A_{n} 
\propto
{\rm tr}\left( T^{i_{1}}\cdots T^{i_{n}} \right) A\left( i_{1},\cdots,i_{n} \right) + 
{\rm permutations}\,,
}
where $\{T^{i}\}$ are the generators of $U\left( N \right)$. The color ordered amplitudes $A\left( i_{1},\cdots,i_{n} \right)$ on the disk are written as an integral over $n-3$ points (three points are fixed by $SL\left( 2,\IR \right)$ invariance) and are characterized by different insertion's positions. As a consequence, deforming the contour of integration allows to write one $A$ in terms of the others, weighted by functions of momenta.
The reality of $A$ implies that the imaginary part of the obtained equation is zero and its real part is equal to the original amplitude. Taking the field theory limit ($\alpha'\rightarrow 0$), the former relation reconstructs the BCJ's, while the latter the KK's. These relations together with the invariance of \stringopenamplgeneral\ under cyclic permutations of the color indices, constrain the number of the $A$'s to $\left( n-3 \right)!$. Notice that this is the minimum number of independent color ordered amplitude \refs{\BjerrumBohrRD,\BjerrumBohrZS}.

BCJ relations can be derived from BRST symmetry by observing the following 
fundamental fact: any ghost-number 2 combination $\Omega^{(2)}$ provides 
a suitable vertex operator satisfying
\eqn\eqqA{
\langle Q \Omega^{(2)} \rangle =0\,,
}
since the BRST charge acts on the invariant vacuum. Given an amplitude with 
$n$ external legs, we can construct several examples of $\Omega^{(2)}$. For example, in the 
case of 4--point we have $V_i T_{ijk}$ and $T_{ij} T_{kl}$ (here we use the notation of Mafra et al. where $V_i$ 
stands for the unintegrated vertex operators and $T_{ij}$ are the residue of the fusion rules between $V_i$ and the integrand of 
the integrated vertex $U_j$); in the case of 5-point 
functions we have $V_i T_{ijkl}$ and $T_{ij} T_{klm}$ and so on. Obviously, increasing the 
number of legs also the number of possible candidates for $\Omega^{(2)}$ and 
that of possible relations increase. In the next sections, we study the relations 
for 4--,  5-- and, finally, 6--point functions. 


\subsec{4-point Functions}

Let us define the building blocks:
\eqn\defnquattro{
n_{ij[kl]} = \langle V_i V_j T_{kl} \rangle \,.
}
Due to their algebraic properties, they obey the following symmetries
\eqn\nquattrosymm{
n_{ji[kl]} = - n_{ij[kl]} \,, \quad
n_{ij[lk]} = - n_{ij[kl]}\,,
}
immediately following from the definition \defnquattro. On the other hand, if we consider the ghost number 2 quantity $\Omega^{(2)}= 
T_{ij} T_{kl}$, the $Q$-exactness implies
\eqn\qexcnquattro{
0 = {1\over s_{ij}} \langle Q \left( T_{ij} T_{kl} \right) \rangle = n_{ij[kl]} - n_{kl[ij]}\,.
}
Therefore, the symmetry of $n$'s under the exchange of the couples of indices  
is not a consequence of the definition of $n$'s, but relies on BRST invariance. Let us choose instead $\Omega^{(2)} = V_i T_{jkl}$: its $Q$-exactness, together with the fact that, for 4-point functions, $s_{ijk} =0$, gives
\eqn\bcjdim{
0 = {1\over s_{jk}} \langle Q \left( V_i T_{jkl} \right) \rangle = - \langle V_i \left( V_j T_{kl} + V_k T_{lj} + V_l T_{jk} \right) \rangle\,,
}
leading to 
\eqn\bcjquattro{
n_{ij[kl]} + n_{ik[lj]} + n_{il[jk]} = 0 \,.
}
This relation is usually called BCJ relation and selects two independent 
$n_{ij[kl]}$ providing a basis for these amplitudes. 


\subsec{5--point Functions}

For the 5--point amplitude the building blocks are
\eqn\ndefncinque{
n_{\left[ ij \right]k\left[ lm \right]}
=
\langle
T_{ij} V_{k} T_{lm}
\rangle
\,,
\quad
m_{\left[ ij \right]klm}=
\langle
V_{i}V_{j}T_{klm}
\rangle\,,
}
notice that there are $15$ independent $n$'s and $15$ independent $m$'s that, by construction, have the following  structural symmetries
\eqn\ncinquestructsym{
n_{\left[ ji \right]k\left[ lm \right]}=
- n_{\left[ ij \right]k \left[ lm \right]} 
\,,\quad
n_{\left[ ij \right]k\left[ ml \right]}=
- n_{\left[ ij \right]k \left[ lm \right] }
\,,\quad
n_{[lm]k[ij]} = - n_{[ij]k[lm]}  
}
$$m_{[ji]klm} = - m_{[ij]klm}\,, \quad m_{[ij]lkm} = - m_{[ij]klm}\,, \quad m_{[ij]klm} + m_{[ij]lmk} + m_{[ij]mkl} = 0.$$
As in the case of 4--point amplitudes, there are 2 $Q$-exactness relations
\eqn\ncinqueQexactnessuno{
{1\over s_{ij}} 
\langle
Q\left( T_{ij}T_{klm} \right)
\rangle = 0\,,
}
\eqn\ncinqueQexactnessdue{
\langle
Q\left( 
V_{m}T_{ijkl}
\right)
\rangle = 0\,,}
which lead to, respectively
\eqn\ncinqueQexactnesstre{
- m_{\left[ ij \right]klm} +
\left( {s_{kl}\over s_{ij}}-1 \right)n_{\left[ ij \right]m\left[ kl \right]}
+ {s_{kl}\over s_{ij}}n_{\left[ ij \right]k\left[ lm \right]}
- {s_{kl}\over s_{ij}}n_{\left[ ij \right]l\left[ mk \right]} = 0
\,,
}
\eqn\BCJncinque{
s_{ijk}n_{\left[ mk \right]l\left[ ij \right]}
+
{s_{ijk}s_{ij} \over s_{mk}}\left( 
n_{\left[ ij \right]l \left[ mk \right]}
+
n_{\left[ jl \right]i \left[  mk\right]}
+
n_{\left[ li \right]j \left[  mk\right]}
\right)+
}
$$
+
s_{ijk}n_{\left[ lm \right]k\left[ ij \right]}
+
s_{ij}\left( 
n_{\left[ ij \right]k \left[  lm\right]}
+
n_{\left[ jk \right]i \left[  lm\right]}
+
n_{\left[ ki \right]j \left[  lm\right]}
\right)+
$$
$$
+
s_{ij}n_{\left[ jk \right]l\left[ mi \right]}
+
{s_{ij}s_{jk} \over s_{mi}}\left( 
n_{\left[ mi \right]l \left[  jk\right]}
+
n_{\left[ mi \right]j \left[  kl\right]}
+
n_{\left[ mi \right]k \left[  lj\right]}
\right)+
$$
$$
+
s_{ij}n_{\left[ mj \right]l\left[ ik \right]}
+
{s_{ij}s_{ki} \over s_{mj}}\left( 
n_{\left[ ik \right]l \left[  mj\right]}
+
n_{\left[ kl \right]i \left[  mj\right]}
+
n_{\left[ li \right]k \left[  mj\right]}
\right)+
$$
$$
+
s_{ij}n_{\left[ ij \right]l\left[ mk \right]}
+
{s_{ij}^{2} \over s_{mk}}\left( 
n_{\left[ mk \right]l \left[  ij\right]}
+
n_{\left[ mk \right]i \left[  jl\right]}
+
n_{\left[ mk \right]j \left[  li\right]}
\right)+
$$
$$
+
\left( s_{ijk}-s_{ij} \right)
n_{\left[ ij \right]m\left[ kl \right]}
+
s_{ij}\left( 
n_{\left[ ik \right]m\left[ jl \right]}
+
n_{\left[ il \right]m\left[ jk \right]}
\right) = 0
\,.
$$

The former relation states that $m$ can always be written as linear combination of $n$'s, while decoupling the latter according to the momentum factor, we determine the $9$ independent BCJ relations
\eqn\ncinqueBCJ{
n_{[ij]k[lm]} + n_{[jk]i[lm]} + n_{[ki]j[lm]} = 0\,.
}
Notice that according to \BjerrumBohrZS, there might be additional solutions to eqs. \BCJncinque\ known as generalized BCJ relations. We have explored this possibility. 


\subsec{6-point Function}

The building blocks needed for the construction of the 6--point amplitude are the following
\eqn\bbsix{
n_{i[jk]lmn} = \langle V_i T_{jk} T_{lmn} \rangle\,,
}
$$m_{[ij]klmn} = \langle V_i V_j T_{klmn} \rangle\,,$$
$$l_{[ij][kl][mn]} = \langle T_{ij} T_{kl} T_{mn}\rangle\,.$$
Now there are 3 different $Q$--exactness conditions 
\eqn\qexsix{
\langle Q V_i T_{jklmn} \rangle = 0\,,
}
$$\langle Q T_{ij} T_{klmn} \rangle = 0\,,$$
$$\langle Q T_{ijk} T_{lmn} \rangle = 0\,,$$
from which we can deduce the following BCJ relations
\eqn\bcjsix{
n_{i[jk]lmn} = n_{n[lm]jki}\,, \quad n_{i[jk]lmn} + n_{j[ki]lmn} + n_{k[ij]lmn} = 0\,,}
$$l_{[ij][kl][mn]} = n_{l[mn]ijk} - n_{k[mn]ijl}\,, $$
$$m_{[mn]ijkl} = n_{l[mn]ijk} \,,$$
$$n_{n[ij]klm} - n_{m[ij]kln} = n_{l[mn]ijk} - n_{k[mn]ijl}\,,$$
that add to the ones given by the structure of the building blocks. Notice that due to the second and third equations in \bbsix\ the $l$ and $m$ blocks are not needed for the construction of the amplitudes.


\newsec{$\tau$'s vs $n$'s}

Here we discuss the relations between Mafra's bulding blocks (BB) and the new objects $\tau$, introduced in \BernIA, which appear to be the natural generalization of the form factors from SYM to Supergravity amplitudes.
For n--point functions there is always a fundamental equation which relates $n$ with $\tau$. This expression reads, for $n=4$,5 and 6:
\eqn\entaufour{n_{ij[kl]} = \tau_{i[j[k,l]]} \equiv \tau_{ijkl} - \tau_{ijlk} - \tau_{iklj} + \tau_{ilkj},
}
\eqn\entaufive{n_{[ij]k[lm]} = \tau_{i[j[k[l,m]]]},
}
\eqn\entausix{
n_{i[jk]lmn} = \tau_{j[k[i[n[l,m]]]]}.
}
It should be noticed that the above definitions are fully compatible with the structural symmetries of the building blocks $n$, provided the $\tau$'s are cyclic. 
To invert the above equations and get  $\tau=\tau(n)$ one can proceed as follows: first express the $\tau$'s in terms of all the BBs
\eqn\taralloA{
\tau_p = \sum_I \alpha^I n^I_p \,,
} 
provided that the cyclic symmetry of $\tau$ is respected. The index $I$ runs over the independent $n$'s and $p$ labels the number of external legs. The symmetries on $\tau$ imply some conditions on the coefficients $\alpha^I$, but in order that the KK are satisfied one needs additional symmetries stemming from eq. $\langle Q \Omega^{(2)}\rangle=0$.  

Notice that for 5-- and higher point functions there are additional elements. Actually, we can define different types of BBs, namely $m_{[ij]klm}$ and $n_{[ij]k[lm]}$ for 5--point and $n_{i[jk]lmn}$, $m_{[ij]klmn}$ and $l_{[ij][kl][mn]}$ for 6--point, as discussed respectively in sec. 2.2 and 2.3. Moreover, in these cases, some existing relations have been derived guaranteeing that the $m$'s (respectively $m$'s and $l$'s) are expressible in terms of $n$'s. Therefore the latter are the only necessary ingredients for the $\tau$'s. However, we can also reverse the argument proposing the following generic decomposition 
\eqn\taralloB{
\tau = \sum_{I,A} \alpha^I_A n^I_A \,, 
}
where $A$ labels the BBs according to their operatorial structure, and $I$ runs over the independent ones.
By imposing the KK relations we found that the complete BCJ relations must be used in order to express the $m$'s and $l$'s in terms of the $n$'s. 

The reverse expression is apparently unique for 4-- and 5--point, and have the form of
\eqn\revexpfour{
\tau_{ijkl} = {1\over 6} (n_{ij[kl]} + n_{kl[ij]})\,, 
}
and
\eqn\revexpfive{
\tau_{ijklm} = {1\over 20} \sum_{{\rm cycle}} (n_{[ij]k[lm]})\,.
}
For 6--point an expression (supposedly not unique) has been guessed in \BernIA.
The $\tau$'s are cyclic on their indices and automatically satisfy KK relations if the $n$'s possess their structural symmetries and satisfy BCJ's as discussed above. 
Moreover, the number of independent $n$'s considering structural symmetries and BCJ's always match the number of independent $\tau$'s surviving the KK relations, that is $(n-2)!$.


\newsec{$6$--point Closed Amplitude}

Here we extend the work presented in \GrassiUF, deriving the 6--point amplitude for closed string.
Following the procedure proposed in \BernIA\ we have
\eqn\closedsixamp{
A_{p}^{closed} 
= i \left( k\over 2 \right)^{p-2}\sum_{\sigma} A^{L}_{p} \left( \sigma \right) \tau^{R} \left( \sigma \right) \,,}
where $\sigma$ labels all non-equivalent permutations of the p indices, $k$ is the gravitational coupling and $A^{L}_{p}$ is a p--point color-ordered open amplitude.

In \BernIA, supergravity amplitudes are found using a set of BBs different from the one by Mafra, namely the $\tau$'s. The amplitude is written formally as the sum over all non-equivalent permutations of the product of $A^{Open}$ and $\tau$, where the former is to be intended left- and the latter right-moving.
Expressing all the amplitudes in terms of $\tau$, it is easy to show that the expression is symmetric in the exchange of left- and right-movers, from which $Q$--invariance is deduced. This is the procedure used in \GrassiUF. 
 In this paper we propose a different way to build the amplitude and to show its $Q$--closure.
Take for instance the 4--point function:
\eqn\fourptsclosed{
A^{Closed} = \sum_\sigma A^{Open}_{L,\sigma} (n^L) \tau^R_\sigma\,,
}
where 
\eqn\fourptsopen{
A^{Open}_{L,ijkl} = {{n^L_{ij[kl]}}\over{s_{kl}}} + {{n^L_{jk[li]}}\over{s_{il}}} + {{n^L_{kl[ij]}}\over{s_{ij}}} + {{n^L_{li[jk]}}\over{s_{jk}}}\,.
}
This expression is manifestly left--BRST invariant: indeed the $\tau^R$ are left invariant (since they are not touched by $Q_L$) and $A^{Open}_{L,\sigma} (n^L)$ are manifestly $Q_L$ invariant:
\eqn\quclos{
Q A_{ijkl}^{Open} = {1\over{s_{kl}}} Q n_{ij[kl]} + {1\over{s_{li}}} Q n_{jk[li]} + {1\over{s_{ij}}} Q n_{kl[ij]} + {1\over{s_{jk}}} Q n_{li[jk]} = 0 \,.
}
Now, we compute the complete expression for $A^{Closed}$ by summing over the 6 independent permutations
\eqn\aclos{
A^{Closed} = \left({{n^L_{ij[kl]}}\over{s_{kl}}} + {{n^L_{jk[li]}}\over{s_{li}}} + {{n^L_{kl[ij]}}\over{s_{ij}}} + {{n^L_{li[jk]}}\over{s_{jk}}} \right) \tau^R_{ijkl} + 
}
$$+ \left({{n^L_{ij[lk]}}\over{s_{lk}}} + {{n^L_{jl[ki]}}\over{s_{ki}}} + {{n^L_{lk[ij]}}\over{s_{ij}}} + {{n^L_{ki[jl]}}\over{s_{jl}}} \right) \tau^R_{ijlk} +$$
$$+ \left({{n^L_{ik[jl]}}\over{s_{jl}}} + {{n^L_{kj[li]}}\over{s_{li}}} + {{n^L_{jl[ik]}}\over{s_{ik}}} + {{n^L_{li[kj]}}\over{s_{kj}}} \right) \tau^R_{ikjl} +$$
$$+ \left({{n^L_{ik[lj]}}\over{s_{lj}}} + {{n^L_{kl[ji]}}\over{s_{ji}}} + {{n^L_{lj[ik]}}\over{s_{ik}}} + {{n^L_{ji[kl]}}\over{s_{kl}}} \right) \tau^R_{iklj} +$$
$$+ \left({{n^L_{il[jk]}}\over{s_{jk}}} + {{n^L_{lj[ki]}}\over{s_{ki}}} + {{n^L_{jk[il]}}\over{s_{il}}} + {{n^L_{ki[lj]}}\over{s_{lj}}} \right) \tau^R_{iljk} +$$
$$\left({{n^L_{il[kj]}}\over{s_{kj}}} + {{n^L_{lk[ji]}}\over{s_{ji}}} + {{n^L_{kj[il]}}\over{s_{il}}} + {{n^L_{ji[lk]}}\over{s_{lk}}} \right) \tau^R_{ilkj}\,.$$
It is easy to see that, collecting the different poles, we end up with expressions characterized by the same $n^L$ in front, multiplying a linear combination of $\tau^R$'s
\eqn\quattrocl{
A^{Closed} = {{n^L_{ij[kl]}} \over {s_{kl}}} \tau^R_{i[j[k,l]]} + 
{{n^L_{ik[lj]}} \over {s_{jl}}} \tau^R_{i[k[l,j]]} + 
{{n^L_{il[jk]}} \over {s_{jk}}} \tau^R_{i[l[j,k]]} +}
$$
+{{n^L_{il[jk]}} \over {s_{il}}} \tau^R_{i[l[j,k]]} + 
{{n^L_{ik[lj]}} \over {s_{ik}}} \tau^R_{i[k[l,j]]} + 
{{n^L_{ij[kl]}} \over {s_{ij}}} \tau^R_{i[j[k,l]]}\,,
$$
then we identify that combination with $n^R$
\eqn\quattrotaun{
n^R_{ij[kl]} \equiv \tau^R_{i[j[k,l]]}
} 
Doing so, we actually require the symmetry in the exchange of left and right. Finally, rewriting the closed string amplitude as
\eqn\aswapped{
A^{Closed} = \sum_\sigma \tau^L_\sigma A^{Open}_{R,\sigma}\,,
}
we can apply the right--BRST charge and, while $\tau^L$ is left untouched by $Q_R$, $A^{Open}_{R,\sigma}$ is manifestly invariant. We can thus conclude that $A^{Closed}$ is $Q$--invariant. Notice that the above combination of $\tau$'s satisfies the BCJ relations, as we explicitly checked. This is a consequence of the request that the $\tau$'s must satisfy KK relations, but a more precise statement is needed: if the $\tau$'s fulfill the KK relations, there are exactly $(n-2)!$ independent building blocks. On the other side, if the $n$'s satisfy the BCJ, they also are expressed in terms of $(n-2)!$ building blocks. Thus, the precise statement is that the relation between $\tau$'s and $n$'s is a one-by-one correspondence if and only if both of them satisfy the respective identities. Swapping between $\tau$'s and $n$'s is allowed only if the number of independent building blocks is the same.

Let us now consider 5--point amplitudes. 
\eqn\openfive{
A^{Open}_{ijklm} = {{n_{[ij]k[lm]}} \over {s_{ij} s_{lm}}} + {{n_{[jk]l[mi]}} \over {s_{jk} s_{im}}} + {{n_{[kl]m[ij]}} \over {s_{kl} s_{ij}}} + {{n_{[lm]i[jk]}} \over {s_{lm} s_{jk}}} + {{n_{[mi]j[kl]}} \over {s_{im} s_{kl}}}\,.}
Applying the same technique the following result for closed amplitude is found
\eqn\fiveclosed{
A^{Closed} = {{n^L_{[ij]k[lm]}} \over {s_{ij} s_{lm}}} \tau^R_{i[j[k[l,m]]]} + 
{{n^L_{[il]k[mj]}} \over {s_{il} s_{jm}}} \tau^R_{i[l[k[m,j]]]} +
{{n^L_{[im]k[jl]}} \over {s_{im} s_{jl}}} \tau^R_{i[m[k[j,l]]]} +}
$$
+ {{n^L_{[jk]l[mi]}} \over {s_{jk} s_{im}}} \tau^R_{j[k[l[m,i]]]} +
{{n^L_{[jm]l[ik]}} \over {s_{jm} s_{ik}}} \tau^R_{j[m[l[i,k]]]} +
{{n^L_{[ji]l[km]}} \over {s_{ij} s_{km}}} \tau^R_{j[i[l[k,m]]]} +
$$
$$
+ {{n^L_{[kl]m[ij]}} \over {s_{kl} s_{ij}}} \tau^R_{k[l[m[i,j]]]} +
{{n^L_{[ki]m[jl]}} \over {s_{ik} s_{jl}}} \tau^R_{k[i[m[j,l]]]} +
{{n^L_{[kj]m[li]}} \over {s_{jk} s_{il}}} \tau^R_{k[j[m[l,i]]]} +
$$
$$
+ {{n^L_{[lm]i[jk]}} \over {s_{lm} s_{jk}}} \tau^R_{l[m[i[j,k]]]} +
{{n^L_{[lj]i[km]}} \over {s_{lj} s_{km}}} \tau^R_{l[j[i[k,m]]]} +
{{n^L_{[lk]i[mj]}} \over {s_{lk} s_{jm}}} \tau^R_{l[k[i[m,j]]]} +
$$
$$
+ {{n^L_{[mi]j[kl]}} \over {s_{mi} s_{kl}}} \tau^R_{m[i[j[k,l]]]} +
{{n^L_{[mk]j[li]}} \over {s_{mk} s_{il}}} \tau^R_{m[k[j[l,i]]]} +
{{n^L_{[ml]j[ik]}} \over {s_{ml} s_{ik}}} \tau^R_{m[l[j[i,k]]]}\,,
$$
from which it is natural to identify
\eqn\ident{
\tau^R_{i[j[k[l,m]]]} \equiv n^R_{[ij]k[lm]}\,,}
and again the closed amplitude is $Q$--invariant by left/right symmetry.

Let us now turn our attention to the case of 6 external legs. First of all we recall the the open string 6--point amplitude proposed by Mafra
\eqn\opsix{
A^{Open}_{ijklmn} = \langle {1 \over 2} {{T_{ijk} T_{lm} V_n}\over{s_{ijk}s_{lm}s_{ij}}} - {1 \over 2} {{T_{ijk} T_{mn} V_l}\over{s_{ijk}s_{mn}s_{ij}}} - {1 \over 2} {{T_{jki} T_{lm} V_n}\over{s_{ijk}s_{lm}s_{jk}}} + {1 \over 2} {{T_{jki} T_{mn} V_l}\over{s_{ijk}s_{mn}s_{jk}}} + } 
$$ + {1 \over 3} {{T_{ij}T_{kl}T_{mn}}\over{s_{ij}s_{kl}s_{mn}}} +  {\rm{cyclic}}(ijklmn) \rangle\,,$$
which in terms of our BBs it becomes
\eqn\ouropsix{
A^{Open}_{ijklmn}
=
{1 \over 2}
{n_{l[mn]ijk}
\over
s_{ijk}s_{ij}s_{mn}
}
+{1 \over 2}
{n_{n[lm]jki}
\over
s_{ijk}s_{jk}s_{lm}
}
-{1 \over 2}
{n_{l[mn]jki}
\over
s_{ijk}s_{jk}s_{mn}
}
-
{1 \over 2}
{n_{n[lm]ijk}
\over
s_{ijk}s_{ij}s_{lm}
}
+}
$$
+{1 \over 3}
{l_{[ij][kl][mn]}
\over
s_{ij}s_{kl}s_{mn}
}
+{\rm cyclic}\left( ijklmn \right)
\,.
$$
Computing \closedsixamp\ and applying the strategy above explained, we find that the $n$'s are always multiplied by the same linear function of $\tau$'s 
\eqn\MYentausix{
n^{L}_{ijklmn}\left( \tau^{R}_{j[k[i[n[l,m]]]]} \right)
\,.
}
Notice that the $l$'s are multiplied by combination of $\tau$'s which is compatible with \bcjsix\ and \MYentausix. 
To have L/R symmetry satisfied, we then define
\eqn\entausix{
n^{R}_{i[jk]lmn} = \tau^{R}_{j[k[i[n[l,m]]]]}\,.
}
The 6--point amplitude for closed string is then closed under the action of $Q=Q^{L}+Q^{R}$.

Let us discuss in more detail the computation which leads to \entausix. 
Starting from \fourptsclosed
\eqn\APPclosed{
A^{Closed} = \sum_\sigma A^{Open}_{L,\sigma} (n^L) \tau^R_\sigma\,,
}
we write explicitly $A^{Open}_{L}$ as in \ouropsix, 
\eqn\ENDentausix{
A^{Closed} = 
\sum_{\sigma} \left( {1 \over 2} {n^{L}_{l[mn]ijk} \over s_{ijk}s_{ij}s_{mn}}
+ {1 \over 2} {n^{L}_{n[lm]jki} \over s_{ijk}s_{jk}s_{lm}}
- {1 \over 2} {n^{L}_{l[mn]jki} \over s_{ijk}s_{jk}s_{mn}}
- {1 \over 2} {n^{L}_{n[lm]ijk} \over s_{ijk}s_{ij}s_{lm}}
+ \right.}
$$
\left. 
+ {1 \over 2} {n^{L}_{m[ni]jkl} \over s_{jkl}s_{jk}s_{ni}}
+ {1 \over 2} {n^{L}_{i[mn]klj} \over s_{jkl}s_{kl}s_{mn}}
- {1 \over 2} {n^{L}_{m[ni]klj} \over s_{jkl}s_{kl}s_{ni}}
- {1 \over 2} {n^{L}_{i[mn]jkl} \over s_{jkl}s_{jk}s_{mn}}
+ \right.
$$
$$
\left.
+ {1 \over 2} {n^{L}_{n[ij]klm} \over s_{klm}s_{kl}s_{ij}}
+ {1 \over 2} {n^{L}_{j[ni]lmk} \over s_{klm}s_{lm}s_{ni}}
- {1 \over 2} {n^{L}_{n[ij]lmk} \over s_{klm}s_{lm}s_{ij}}
- {1 \over 2} {n^{L}_{j[ni]klm} \over s_{klm}s_{kl}s_{ni}}
+ \right.
$$
$$
\left.
+ {1 \over 2} {n^{L}_{i[jk]lmn} \over s_{lmn}s_{lm}s_{jk}}
+ {1 \over 2} {n^{L}_{k[ij]mnl} \over s_{lmn}s_{mn}s_{ij}}
- {1 \over 2} {n^{L}_{i[jk]mnl} \over s_{lmn}s_{mn}s_{jk}}
- {1 \over 2} {n^{L}_{k[ij]lmn} \over s_{lmn}s_{lm}s_{ij}}
+ \right.
$$
$$
\left.
+ {1 \over 2} {n^{L}_{j[kl]mni} \over s_{mni}s_{mn}s_{kl}}
+ {1 \over 2} {n^{L}_{l[jk]nim} \over s_{mni}s_{ni}s_{jk}}
- {1 \over 2} {n^{L}_{j[kl]nim} \over s_{mni}s_{ni}s_{kl}}
- {1 \over 2} {n^{L}_{l[jk]mni} \over s_{mni}s_{mn}s_{jk}}
+ \right.
$$
$$
\left.
+ {1 \over 2} {n^{L}_{k[lm]nij} \over s_{nij}s_{ni}s_{lm}}
+ {1 \over 2} {n^{L}_{m[kl]ijn} \over s_{nij}s_{ij}s_{kl}}
- {1 \over 2} {n^{L}_{k[lm]ijn} \over s_{nij}s_{ij}s_{lm}}
- {1 \over 2} {n^{L}_{m[kl]nij} \over s_{nij}s_{ni}s_{kl}}
+ \right.
$$
$$
\left.
+ {1 \over 3} {l^L_{[ij][kl][mn]} \over s_{ij}s_{kl}s_{mn}}
+ {1 \over 3} {l^L_{[jk][lm][ni]} \over s_{jk}s_{lm}s_{ni}}
+ {1 \over 3} {l^L_{[kl][mn][ij]} \over s_{kl}s_{mn}s_{ij}}
+ \right.
$$
$$
\left.
+ {1 \over 3} {l^L_{[lm][ni][jk]} \over s_{lm}s_{ni}s_{jk}}
+ {1 \over 3} {l^L_{[mn][ij][kl]} \over s_{mn}s_{ij}s_{kl}}
+ {1 \over 3} {l^L_{[ni][jk][lm]} \over s_{ni}s_{jk}s_{lm}}
\right) \, \tau^R_{ijklmn}\,.
$$
Using \bcjsix\ to write $l$ as function of $n$ and summing over all the $120$ inequivalent permutations $\sigma$ of the $6$ indices ({\it i.e.} fixing for example the first index and permutating the others) we obtain the following handy expression
\eqn\MYentausix{
A^{Closed} =
{1\over 2}\sum_{\beta}{ 1\over s_{lmn}s_{lm}s_{jk}} n^L_{i[jk]lmn} \left( \tau^R_{j[k[i[n[l,m]]]]} \right) + }
$$
+2 \sum_{\beta}{ 1\over s_{ij}s_{kl}s_{mn}}
\left( n^L_{l[mn]ijk} - n^L_{k[mn]ijl} \right)
\left( \tau^R_{m[n[l[k[i,j]]]]} - \tau^R_{m[n[k[l[i,j]]]]} \right) \,,
$$
where $\beta$ is the set of the different strings (made with $i,j,k,l,m,n$ indices) which gives rise to independent structures in the poles. For 6 external legs, without considering momentum conservation, there are $180$ independent pole structures for $n$ and $15$ for $l$. 

As we have already said, to have $Q$-closure we impose
\eqn\entausixAP{
n^R_{i[jk]lmn} = \tau^R_{j[k[i[n[l,m]]]]}\,,
}
and with this definition we write the the 6--point closed amplitude
\eqn\MYentausixfin{
A^{Closed} =
{1\over 2}\sum_{\beta}{ 1\over s_{lmn}s_{lm}s_{jk}} n^{L}_{i[jk]lmn} n^{R}_{i[jk]lmn}
+}
$$
+ 2 \sum_{\beta} { 1\over s_{ij}s_{kl}s_{mn}}
\left( n^{L}_{l[mn]ijk} - n^{L}_{k[mn]ijl} \right)
\left( n^{R}_{l[mn]ijk} - n^{R}_{k[mn]ijl} \right) \,,
$$
which is clearly L/R symmetric and thus $Q$-invariant.

\newsec{Conclusions}

We discuss the relations between different building blocks emerging in the construction of 
SYM and supergravity amplitudes. We discuss the role of the KK and of the BCJ relations and 
how they can be derived from the BRST symmetry. Finally, we discuss the decomposition of the 
of the 6-point supergravity amplitude and its BRST invariance. A future endeavor will be the 
comparison of the final expression with the tree level amplitudes in the literature. That amounts 
to compute the pure spinor correlation functions explicitly and it will be published somewhere else. 

\vskip .5cm
\hskip -.7cm  {\bf Acknowledgments}
\vskip .5cm

We thank C. Mafra and P. Vanhove for useful discussions. 


\appendix{A}{${\cal{B}}$ algebra}

As discussed in \MafraJQ, instead of the buliding blocks $V_i, T_{ij}, T_{ijk}, ..... $ it is sometimes convenient 
to use the new objects ${\cal M}_{i}, {\cal M}_{ij}, {\cal M}_{ijk}, ......$ which are related to the first ones by simple redefinitions and 
linear combinations. The advantage of this translation is the simplification of the BRST transformation rules which become 
\eqn\BRSTMM{
Q {\cal M}_{i} = 0\,, \quad
Q {\cal M}_{ij} = {\cal M}_{i} {\cal M}_{j}\,, \quad
Q {\cal M}_{ijk} = {\cal M}_{ij} {\cal M}_{k} + {\cal M}_{i} {\cal M}_{jk} \,, \quad
\dots
}
In order to deal with the complete set of these equations it is convenient to introduce 
the follwing combination: given auxiliary variables $\xi^i$ which are multiplied tensorially as 
$\xi^i \otimes \xi^j \otimes \xi^k \dots$, we can introduce the 
field 
\eqn\newfield{
\Phi = \xi^i {\cal M}_i + \xi^i \otimes \xi^j {\cal M}_{ij} + \xi^i \otimes \xi^j \otimes \xi^k \, {\cal M}_{ijk} + \dots\,.
 }
Assuming that the new variables are inert under the BRST transformation rules, 
we have 
\eqn\newEQ{
Q \Phi = \Phi \otimes \Phi\,.
}
which has a very suggestive form. It can be easily verified that, by picking each term from both side 
of the equation with the same power of $\xi$'s, one reproduces eqs. \BRSTMM.  Notice that the field 
$\Phi$ is anticommuting (since we assigned bosonic statistic to $\xi$'s) and therefore it is easy to 
check the nilpotency of $Q$ consistently with the above equation. 

At this point, one can introduce a new 
operator ${\cal B}$ which has the property 
\eqn\commB{
\{{\cal B}, Q \} = (1 - T^1_{\xi=0}) \,,
}
where $ T^1_{\xi=0}$ is the Taylor expansion of a polynomial of $\xi$'s around $\xi= 0$. The second term 
on the r.h.s. it is needed to show that the BRST on the space of the building blocks has no cohomology. 
Acting directly with the operator ${\cal B}$ on the tensor product of two fields $\Phi$, it yields 
\eqn\actionB
{
{\cal B} (\Phi \otimes \Phi) = {\cal B} Q  \Phi = (1 - T_{\xi=0})\Phi - Q {\cal B}\Phi \,,
}
and, imposing the choice ${\cal B} \Phi = 0$, we get 
\eqn\actionBB{
{\cal B} (\Phi \otimes \Phi) = (1 - T_{\xi=0})\Phi\,.
} 
The operator ${\cal B}$ is not a derivation, otherwise the above equation would have implied that 
the r.h.s. vanishes. By expanding the above equation, 
we get the interesting results 
\eqn\actionBBB{
{\cal B}({\cal M}_i {\cal M}_j) = {\cal M}_{ij}\,,\quad
{\cal B}({\cal M}_{ij} {\cal M}_k + {\cal M}_{i} {\cal M}_{jk}) = {\cal M}_{ijk}\,,}
$$
{\cal B}({\cal M}_{ijk} {\cal M}_l + {\cal M}_{ij} {\cal M}_{kl} + {\cal M}_{i} {\cal M}_{jkl}) = {\cal M}_{ijkl}\,,\quad
\dots 
$$
It is still obscure how the new operator ${\cal B}$ should act on tensor product of 3 or more $\Phi$. For example we may have that 
\eqn\newacA{
{\cal B}(\Phi \otimes \Phi \otimes \Phi) = (1 - T_{\xi=0})\Phi \otimes \Phi - \Phi \otimes (1 - T_{\xi=0})\Phi\,,
}
where the operator ${\cal B}$ has been supposed to act on consequent pairs of fields $\Phi$. 
This is only a preliminary analysis, but in our opinion there should be an operator ${\cal B}$ 
acting on the algebra on building blocks implementing the famous $B$-field.

\listrefs
\bye